\documentstyle[12pt]{article}

\catcode`\@=11
\long\def\@makefntext#1{
\protect\noindent \hbox to 3.2pt {\hskip-.9pt  
$^{{\ninerm\@thefnmark}}$\hfil}#1\hfill}		%CAN BE USED 

\def\@makefnmark{\hbox to 0pt{$^{\@thefnmark}$\hss}}  %ORIGINAL 
	
\def\ps@myheadings{\let\@mkboth\@gobbletwo
\def\@oddhead{\hbox{}
\rightmark\hfil\ninerm\thepage}   
\def\@oddfoot{}\def\@evenhead{\ninerm\thepage\hfil
\leftmark\hbox{}}\def\@evenfoot{}
\def\sectionmark##1{}\def\subsectionmark##1{}}

\renewcommand{\thefootnote}{\fnsymbol{footnote}}

\newcounter{sectionc}\newcounter{subsectionc}\newcounter{subsubsectionc}
\renewcommand{\section}[1] {\vspace*{0.6cm}\addtocounter{sectionc}{1} 
\setcounter{subsectionc}{0}\setcounter{subsubsectionc}{0}\noindent 
	{\normalsize\bf\thesectionc. #1}\par\vspace*{0.4cm}}
\renewcommand{\subsection}[1] {\vspace*{0.6cm}\addtocounter{subsectionc}{1} 
	\setcounter{subsubsectionc}{0}\noindent 
	{\normalsize\it\thesectionc.\thesubsectionc. #1}\par\vspace*{0.4cm}}
\renewcommand{\subsubsection}[1]
{\vspace*{0.6cm}\addtocounter{subsubsectionc}{1}
	\noindent {\normalsize\rm\thesectionc.\thesubsectionc.\thesubsubsectionc. 
	#1}\par\vspace*{0.4cm}}

\newcounter{appendixc}
\newcounter{subappendixc}[appendixc]
\newcounter{subsubappendixc}[subappendixc]

\renewcommand{\appendix}[1] {\vspace*{0.6cm}
        \refstepcounter{appendixc}
        \setcounter{figure}{0}
        \setcounter{table}{0}
        \setcounter{equation}{0}
        \renewcommand{\thefigure}{\Alph{appendixc}.\arabic{figure}}
        \renewcommand{\thetable}{\Alph{appendixc}.\arabic{table}}
        \renewcommand{\theappendixc}{\Alph{appendixc}}
        \renewcommand{\theequation}{\Alph{appendixc}.\arabic{equation}}
        \noindent{\bf Appendix \theappendixc #1}\par\vspace*{0.4cm}}

\def\abstracts#1{{
	\centering{\begin{minipage}{12.2truecm}\footnotesize\baselineskip=12pt\noindent
	\centerline{\footnotesize ABSTRACT}\vspace*{0.3cm}
	\parindent=0pt #1
	\end{minipage}}\par}} 

\renewenvironment{thebibliography}[1]
	{\begin{list}{\arabic{enumi}.}
	{\usecounter{enumi}\setlength{\parsep}{0pt}
\setlength{\leftmargin 1.25cm}{\rightmargin 0pt}
	 \setlength{\itemsep}{0pt} \settowidth
	{\labelwidth}{#1.}\sloppy}}{\end{list}}

\newcounter{itemlistc}
\newcounter{romanlistc}
\newcounter{alphlistc}
\newcounter{arabiclistc}

\newcommand{\fcaption}[1]{
        \refstepcounter{figure}
        \setbox\@tempboxa = \hbox{\footnotesize Fig.~\thefigure. #1}
        \ifdim \wd\@tempboxa > 6in
           {\begin{center}
        \parbox{6in}{\footnotesize\baselineskip=12pt Fig.~\thefigure. #1}
            \end{center}}
        \else
             {\begin{center}
             {\footnotesize Fig.~\thefigure. #1}
              \end{center}}
        \fi}

\newcommand{\tcaption}[1]{
        \refstepcounter{table}
        \setbox\@tempboxa = \hbox{\footnotesize Table~\thetable. #1}
        \ifdim \wd\@tempboxa > 6in
           {\begin{center}
        \parbox{6in}{\footnotesize\baselineskip=12pt Table~\thetable. #1}
            \end{center}}
        \else
             {\begin{center}
             {\footnotesize Table~\thetable. #1}
              \end{center}}
        \fi}

\def\@citex[#1]#2{\if@filesw\immediate\write\@auxout
	{\string\citation{#2}}\fi
\def\@citea{}\@cite{\@for\@citeb:=#2\do
	{\@citea\def\@citea{,}\@ifundefined
	{b@\@citeb}{{\bf ?}\@warning
	{Citation `\@citeb' on page \thepage \space undefined}}
	{\csname b@\@citeb\endcsname}}}{#1}}

\newif\if@cghi
\def\cite{\@cghitrue\@ifnextchar [{\@tempswatrue
	\@citex}{\@tempswafalse\@citex[]}}
\def\citelow{\@cghifalse\@ifnextchar [{\@tempswatrue
	\@citex}{\@tempswafalse\@citex[]}}
\def\@cite#1#2{{$\null^{#1}$\if@tempswa\typeout
	{IJCGA warning: optional citation argument 
	ignored: `#2'} \fi}}

 1
 1
 1

\font\ninerm=cmr9

\newcommand{\be}[1]{\begin{equation}\label{#1}}
\newcommand{\beq}{\begin{equation}}
\newcommand{\ee}{\end{equation}}
\newcommand{\beqn}[1]{\begin{eqnarray}\label{#1}}
\newcommand{\eeqn}{\end{eqnarray}}
\newcommand{\bd}{\begin{displaymath}}
\newcommand{\ed}{\end{displaymath}}
\newcommand{\mat}[4]{\left(\begin{array}{cc}{#1}&{#2}\\{#3}&{#4}\end{array}
\right)}

\def\tanb{\tan\beta}

\def\al{\alpha}

\def\la{\lambda}
\def\La{\Lambda}
\newcommand{\eps}{\varepsilon}

\newcommand{\ov}{\overline}
\renewcommand{\to}{\rightarrow}

\begin{document}

\begin{flushright}
INFNFE 02-97 \\ 
February 1997 
\end{flushright}
\vspace{3mm}

\centerline{\normalsize\bf 2/3 SPLITTING IN SUSY GUT --- HIGGS AS 
GOLDSTONE BOSON} 
\baselineskip=22pt
%\centerline{\normalsize\bf GUIDELINES FOR TYPESETTING A CAMERA-READY}
%\baselineskip=16pt
%\centerline{\normalsize\bf MANUSCRIPT BY COMPUTER}
%\centerline{\footnotesize\sf (For subsequent 20\% photoreduction
%to 17.8 $\times$ 11.9 cm text area)\footnote{The \LaTeX\ source
%file for this document may be used as a template for your
%article, and can be requested by e-mailing {\sf
%worldscp@singnet.com.sg}.}}

%\vfill
%\vspace*{0.6cm}
\centerline{\footnotesize Z. BEREZHIANI}
\baselineskip=13pt
\centerline{\footnotesize\it INFN Sezione di Ferrara, I-44100 Ferrara, 
Italy, and } 
\baselineskip=12pt
\centerline{\footnotesize\it Institute of Physics, Georgian Academy 
of Sciences, 38077 Tbilisi, Georgia} 
\centerline{\footnotesize E-mail: berezhiani@axpfe1.fe.infn.it}
%\vspace*{0.3cm}
%\centerline{\footnotesize and}
%\vspace*{0.3cm}
%\centerline{\footnotesize SECOND AUTHOR'S NAME}
%\baselineskip=13pt
%\centerline{\footnotesize\it Group, Company, Address, City, State ZIP/Zone,
%Country}

%\vfill
\vspace*{0.9cm}
\abstracts{ We briefly review 
the GIFT (Goltstones instead of Fine Tuning) mechanism which provides 
a promising solution to the doublet-triplet splitting problem  
and some other puzzles of the supersymmetric grand unification 
as are the $\mu$-problem, the fermion mass problem, etc.    
It can be naturally implemented by extending the minimal SUSY 
$SU(5)$ model to the gauge $SU(6)$ theory. 
%which symmetry first breaks down to $SU(5)$ and then .  
%The recent status of discussed 
}
 
%\vspace*{0.6cm}
\normalsize\baselineskip=15pt
\setcounter{footnote}{0}
\renewcommand{\thefootnote}{\alph{footnote}}

\section{Introduction}

The concepts of supersymmetry (SUSY) and grand unification theories 
(GUT) constitute the most promising ideas beyond the standard model. 
The present data on $\al_3(M_Z)$ and $\sin^2\theta_W(M_Z)$ are in a 
remarkable agreement with the prediction of the SUSY $SU(5)$ theory 
while exclude the non-supersymmetric $SU(5)$.\cite{Amaldi}   
On the other hand, the minimal supersymmetric standard model (MSSM) 
without GUT is also not in best shape: 
unification at the string scale gives too small $\sin^2\theta_W(M_Z)$. 
In the MSSM the running gauge constants $g_3$, $g_2$ and $g_1$ 
given at $\mu=M_Z$ within the present experimental error bars, 
in their evolution to higher energies join at the scale 
$M_5\simeq 10^{16}$ GeV.\cite{Amaldi}  
Hence, at this scale the $SU(3)\times SU(2)\times U(1)$ symmetry 
can be consistently embedded into $SU(5)$. An additional hint is 
provided by the fact of the $b-\tau$ Yukawa unification at GUT 
energies\cite{BEGN}.  
%which at larger scales can be further extended to 
%or some larger groups, e.g. $SO(10)$, $SU(6)$ or $E_6$.  
%
All these suggest a following paradigm: 

At the scale $M_{\rm str}=g_{\rm str}M_P$ a basic `string' theory 
reduces to a SUSY GUT given by 
a gauge group $G$ ($G=SU(5),SO(10), SU(6)$ etc.) with a gauge 
constant defined by condition $k_Gg_G^2=g_{\rm str}^2$, where 
$k_G$ is a corresponding Kac-Moody level and 
$M_P\simeq 2\cdot 10^{18}$ GeV is a reduced Planck mass. 
At some scale $M_G\geq M_5$ the gauge group breaks down 
to $SU(5)$ subgroup, which then at $M_5\simeq 10^{16}$ GeV
breaks down to the $SU(3)\times SU(2)\times U(1)$. 
The value of the gauge constant at this scale is $g_5\simeq 0.7$. 
%
%It can be further extended to larger symmetry ($G=SO(10), 
%SU(6)$, or $E_6$) at some scale $M_G\geq M_X$. 
In order to keep the gauge coupling 
in the perturbative regime up to the string scale, 
there should not be too many extra states with masses in 
the range from $M_5$ to $M_P$,   
%otherwise it cannot be embeddable in string theory.  

%\footnote{
%Without knowing exactly to what is the basic scale of the theory, 
%in the following under the Planck scale we imply a broad range 
%$M_P\sim 10^{17-19}$ GeV, unless it is specified. In particular,  
%this can be the Planck mass  $M_{Pl}\simeq 10^{19}$ GeV itself, 
%reduced Planck mass or a string scale $\sim 3\cdot 10^{17}$ GeV. }

Below the scale $M_5$ the theory is just MSSM 
%there is a {\em Great Desert} with no extra 
%particles  besides the known ones and their superpartners. 
containing the $SU(3)\times SU(2)\times U(1)$ gauge superfields,  
the chiral superfields of quarks and leptons 
$q_i$, $l_i$, $u^c_i$, $d^c_i$, $e^c_i$ ($i=1,2,3$), and two 
Higgs doublets $\phi_{1,2}$.\footnote{
In other words, no GUT debris is allowed below the scale $M_G$ 
%In other words, no additional states are allowed 
%at intermediate scales between $M_Z$ and $M_G$, 
except the singlets. 
% is disfavoured by the gauge coupling 
%unification and the bottom-tau Yukawa unification constraints.  
The extra complete degenerate $SU(5)$ supemultiplets 
populating the intermediate scales would not affect the 
unification of the gauge constants. However, they would 
contribute the renormalization group (RG) running of the 
Yukawa constants from $M_5$ down to $M_Z$ and thus affect 
the $b-\tau$ Yukawa unification. } 
Barring the gauge sector, its effective Lagrangian is given 
by the superpotential terms: 
\beqn{MSSM} 
& W_{\rm Yuk}= \la^u_{ij} q_iu^c_j\phi_2 + 
\la^d_{ij} q_id^c_j\phi_1 + \la^e_{ij} l_ie^c_j\phi_1, \nonumber \\ 
& W_{\rm Higgs} = \mu \phi_1\phi_2 
\eeqn 
and the soft supersymmetry breaking (SSB) terms 
(tilde labels sfermions): 
\beqn{SSB} 
& L_A = A^u_{ij}\tilde{q}_i\tilde{u}^c_j\phi_2 + 
A^d_{ij}\tilde{q}_i\tilde{d}^c_j\phi_1 + 
A^e_{ij}\tilde{l}_i\tilde{e}^c_j\phi_1 + h.c. , \nonumber \\ 
& L_{\rm sf} = \sum m^2_{ij}\tilde{f}_i^\dagger\tilde{f}_j, 
~~~~~~ f=q,u^c,d^c,l,e^c  \nonumber \\ 
& L_{\rm Higgs} = m^2_1|\phi_1|^2 + m^2_2|\phi_2|^2 + 
(B_\mu \phi_1\phi_2 + h.c.) 
\eeqn 
with masses and dimensional constants of the order of 
supersymmetry breaking scale $m_S\sim m_Z$ (in the 
supergarvity folklore --- the gravitino mass $m_{3/2}$). 
The VEVs 
$\langle \phi_1^0 \rangle = v_1=v\cos\beta$ and 
$\langle \phi_2^0 \rangle = v_2=v\sin\beta$, $v=174$ GeV,  
break the electroweak symmetry and induce the fermion masses. 
%Therefore, the complete Higgs potential of scalars $\phi_{1,2}$ 
%has a form: 
%\be{phi}
%{\cal V}(\phi_1,\phi_2)= (\mu^2 + m_1^2)|\phi_1|^2 + 
%(\mu^2 + m_2^2)|\phi_2|^2 + (B_\mu \phi_1\phi_2 + h.c.) 
%+ (g_2^2 + g_1^2)(|\phi_1|^2 - |\phi_2|^2)^2 
%\ee

%\subsection{Grand DT Hierarchy Problem and small $\mu$ Problem }
%\vspace{-0.35cm}

The softly broken SUSY is the only 
plausible idea that can support the GUT 
against the gauge hierarchy problem\cite{Maiani}. 
%A realistic SUSY GUT should be capable to solve naturally the gauge 
%hierarchy problem. 
At the level of the standard model this is essentially a 
problem of the Higgs mass stability against radiative corrections 
(quadratic divergences). 
It is removed as soon as one appeals to SUSY, which links the 
scalar masses to those of their fermion superpartners while 
the latter are protected by the chiral symmetry. 
In the context of grand unification the gauge hierarchy 
problem concerns rather the origin of scales: why the weak scale 
$v\sim M_W$ is so small as compared to the GUT scale 
$M_5$, which in itself is not far from the Planck scale $M_P$.  
This question is inevitably connected with the 
doublet-triplet (2/3) splitting puzzle\cite{Maiani}: the 
Higgs doublets $\phi_1,\phi_2$ embedded in the GUT multiplets are 
accompanied by the colour triplet partners $\bar T,T$. 
The latter would mediate unacceptably fast proton decay  
(via $d=5$ operators\cite{dim5}) unless they are superheavy, 
with mass $\sim M_5$.  

Yet another problem is a so-called $\mu$-problem\cite{mu}. 
The supersymmetric term $\mu\phi_1\phi_2$ should be small,  
and at first sight $\mu=0$ could be a most natural possibility: 
then the Higgs masses would  emerge entirely from the soft SUSY 
breaking terms with $m_S\sim v$. 
However, $\mu=0$ is excluded experimentally  
and it is required to be  $\sim v$ as well.  
%in SUSY GUTs the gauge hierarchy problem 
%turns into a problem of small (but {\em not} vanishingly small) 
%$\mu$-term: $\mu\sim m_S$. 
Therefore, one has to explain in a natural way, 
why the value of the supersymmetric mass $\mu$ is of the same 
order as that of the soft SUSY breaking mass $m_S$. 

In the context of the realistic SUSY GUTs the 2/3 splitting and 
$\mu$ problems are intimately related to each other and to the 
other key problems as are the origin of the supersymmetry breaking, 
the fermion mass problem, proton stability, etc.

\section{2/3 splitting in SUSY SU(5) } 

In the minimal SUSY $SU(5)$ model the Higgs sector consists of 
the chiral superfields in adjoint representation $\Sigma\sim 24$  
and fundamental representations $H\sim 5$, $\bar H\sim \bar5$: 
$H=(T+\phi_2)$ and $\bar H = (\bar T + \phi_1)$, where  
$\phi_{1,2}$ are the MSSM Higgs doublets and $\bar T,T$ 
colour triplet partners. 
The most general superpotential terms involving these fields 
are the following:
%\footnote{ The Higgs and fermion superfields are distiguished by 
%a matter parity, positive for Higgses and negative for fermions.}  
\be{W-su5}
{\cal W}_{\rm Higgs}= M\Sigma^2 + h\Sigma^3 + 
M_H \bar H H  +  f\bar H \Sigma H 
\ee 
The $SU(5)$ symmetry breaking down to 
$SU(3)\times SU(2)\times U(1)$ is provided by supersymmetric 
ground state $\langle \Sigma \rangle = (2M/3h)\mbox{diag}(2,2,2,-3,-3)$, 
$\langle \bar H, H \rangle =0$. In this vacuum 
the masses of the $T$ and $\phi$ fragments are different: 
$M_3 T\bar T$ and $\mu' \phi_1\phi_2$, where 
$M_3=M_H -(3f/h)M$ and $\mu'=M_H + (2f/h)M$. 
So, the light doublet ($\mu\sim M_W$) versus heavy triplet 
($M_3\sim M_X$) requires  that $hM_H\approx -2fM$, with the accuracy 
of about $10^{-14}$.  Supersymmetry renders this constraint stable 
against radiative corrections.  
However, such a {\it technical solution}\cite{Maiani} is nothing but 
the {\it Fine Tuning} (FT) of the parameters in the superpotential. 

The actual puzzle is to achieve the 2/3 splitting in a natural way, 
without FT. Several attempts have been done, which we briefly 
describe below: 

$\bullet$ {\it Missing doublet mechanism (MDM)}.\cite{MDM}  
In this scenario the 
24-plet $\Sigma$ is replaced by the 75-plet $\Phi$ of $SU(5)$, 
and additional heavy superfields $\Psi\sim 50$ and  
$\bar\Psi\sim \ov{50}$ are introduced. The latter have a bare mass 
term $M_\Psi \bar\Psi \Psi$,  $M_\Psi\sim M_X$, while $M_H H\bar H$ 
%can be also suppressed by some symmetry. 
% In this case the term $\bar H \Phi H$ 
is absend in the superpotential. Instead 
the superpotential contains the terms $\bar H \Phi \Psi$ 
and $H \Phi \bar\Psi$. Then the triplet components $T,\bar T$ 
in $H,\bar H$ can get $\sim M_X$ mass via their mixing to 
triplet states in $\Psi,\bar\Psi$, while the doublets $\phi_{1,2}$, 
will remain massless since 50-plet does not contain doublet fragment. 
The MDM can be motivated by some additional symmetry reasons, 
among which the anomalous $U(1)$ gauge symmetry seems to 
be a most compelling variant\cite{IMDM}. 

%However, the solution of the $\mu$-problem remains open. 
%The small $\mu$-term can appear e.g. from the superpotential term 
%$\mu H\bar H$

New possibilities can emerge with introducing 
%along with $\Sigma$ 
the extra singlet superfield $I$. Then the most general 
renormalizable Higgs superpotential reads: 
\be{W-I} 
{\cal W}_{\rm Higgs}= M\Sigma^2 + h\Sigma^3 + \La^2 I + M'I^2 
+ h'I^3 + h''I\Sigma^2 + M_H \bar{H}H + f\bar{H}\Sigma H 
+ f'I\bar{H}H
\ee 
The 2/3 splitting can be achieved under certain constraints on 
the coupling constants:  

$\bullet$ {\it Sliding singlet mechanism}\cite{sliding} implies 
that $I$ has no couplings in the superpotential except the last term 
in (\ref{W-I}). Then at tree level, with taking into account 
also the SSB terms, the VEV of $I$ can be adjusted to cancell the 
large contributions to the doublet mass from other terms. 
This mechanism, however, was shown to be unstable against radiative 
corrections\cite{no-slid}  -- 
the tadpole diagrams induce the doublet mass $\sim \sqrt{m_S M_G}$. 
%and we will not consider it here. 

$\bullet$ {\it Missing VEV mechanism (MVM)}. One can assume that the 
$\Sigma$ and $I$ together form a {\it reducible}  
representation $\Sigma' \sim 25 = 24+1$ 
and the superpotential has a form: 
\be{W-MVM} 
{\cal W}_{\rm Higgs}= M\Sigma'^2 + h\Sigma'^3 + f\bar{H}\Sigma'H, 
~~~~ (\Sigma')^\al_\beta=\Sigma^\al_\beta + 
\frac{1}{\sqrt{5}} I \delta^\al_\beta 
%M_H\bar{H}H   
\ee 
Since $\Sigma'$ is not traceless, its VEV can be chosen as 
$\langle\Sigma' \rangle=(2M/3h){\rm diag}(1,1,1,0,0)$. 
In this case the triplets get mass $M_3=(2f/3h)M$ while the doublets 
remain massless. 
 
In the $SU(5)$ framework it is difficult to justify 
such a situation, which in  
terms of the general expression (\ref{W-I}) is equivalent to 
the following 5 conditions of the FT: 
\be{FT-1}
M'=M, ~~~~ h'=\frac13 h''= \frac{1}{\sqrt5}h, ~~~~ \La=M_H=0  
\ee 
%and we have assumed also that $\La,M_H=0$.   

Nevertheless, it can be realized in a natural way in models 
with a larger gauge symmetry where both 24 and 1 can be embbedded 
in one irreducible representation which is not constrained to be 
traceless. 
In particular, in $SO(10)$ model  
the 45-plet Higgs of $SO(10)$ containing both 24-plet and 
singlet fragments: $45=1+24+10+\ov{10}$, 
can have a VEV of the form\cite{DiWi}  
%\be{BL-form} 
$\langle A \rangle \sim {\rm diag}(1,1,1,0,0)\otimes i\sigma$.  
%\sigma=\mat{0}{-1}{1}{0}
%\ee
Since 45-plet is antisymmetric, the coupling $H A H$ is not allowed. 
However, one can introduce the additional 10-plet $H'$ with large 
mass term $M'H'H'$ and mix with $H$ through $A$. 

$\bullet$ {\it Goldstones instead of Fine Tuning (GIFT) 
mechanism}.\cite{Inoue,Anselm} 
The superfields $\Sigma$, $I$, $H$ and $\bar H$ can be combined in 
the {\it irreducible} representation $\Sigma'\sim 35$ 
of the $SU(6)$ group: $35=24+1+5+\bar5$. 
%$\Sigma'=\Sigma + (1/\sqrt2)\la_{35}I + \la^+_{6i}\bar H^i 
%+ \la^+_{6i} H_i$. 
Therefore, if the Higgs superpotential 
is a function of the combination $\Sigma'$: 
\be{GIFT}
{\cal W}_{\rm Higgs}= M\Sigma'^2 + h\Sigma'^3 ,  ~~~~~~\Sigma'=
\mat{-\frac{5}{\sqrt{30}}I}{H}{\bar H}{\Sigma+\frac{1}{\sqrt{30}}I} 
\ee 
then it possess a global symmetry $SU(6)$, 
larger than the gauge $SU(5)$ symmetry. 
%In the limit of exact supersymmetry 
The vacuum state can be chosen as 
$\langle\Sigma'\rangle= (M/h){\rm diag}(1,1,1,1,-2,-2)$, 
which breaks the global $SU(6)$ symmetry down to 
$SU(4)\times SU(2)\times U(1)$ subgroup, while its gauged part 
$SU(5)$ breaks down to  standard $SU(3)\times SU(2)\times U(1)$.
The corresponding Goldstone modes are in 
representations $(4,\bar2)+(\bar4,2)$. 
Among those, the $SU(3)$ triplet fragments  $(3,\bar2)+(\bar3,2)$ 
are eaten up by the $SU(5)$ gauge supefields ($X,Y$ bosons) due to 
Higgs mecghanism. As for the fragments 
$(1,\bar2)+(1,2)$ they remain massless as Goldstone modes, 
and can be identified as the MSSM Higgs doublets $\phi_{1,2}$

In view of the general expression (\ref{W-I}) 
this situation requires 9 FT constraints: 
\be{FT-9}
\La=0, ~~~ M=M'=\frac12 M_H, ~~~ 
h=-\sqrt{\frac{15}{8}}h'=\sqrt{\frac{10}{3}}h''=\frac13f=
-\sqrt{\frac{5}{6}}f' 
\ee
However, it can naturally emerge if one extends the 
gauge symmetry to $SU(6)$.\cite{BD} 
One can assume that at the scale $M_6\gg M_5$ the 
local $SU(6)$ symmetry is broken down to $SU(5)$ by some 
Higgses which do not couple $\Sigma\sim 35$ 
Higgses in the Higgs superpotential. Then the superpotential 
of $\Sigma$ would not `feel' that the gauge symmetry is already 
reduced to $SU(5)$ and maintain the global $SU(6)$ invariance.

\section{ SUSY SU(6) Model }
%\vspace{-0.7cm}
%\subsection{Higgs Doublets as Pseudo-Goldstone Bosons }
%\vspace{-0.35cm}

%SUSY $SU(6)$ model\cite{BD} (see also\cite{BDM,BDSBH,PLB,BCL}) 
%was originally designed for the natural solution to the gauge hierarchy 
%and 2/3 splitting problems via the elegant 
%GIFT mechanism\footnote{The 
%GIFT mechanism for the DT splitting was first suggested
%in the context of SUSY $SU(5)$ by assuming an {\em ad hoc} $SU(6)$ 
%global symmetry of the Higgs superpotential\cite{Inoue,Anselm}. 
%Results for fermion masses, however, are specific of the gauged 
%$SU(6)$ theory. }. 
The $SU(6)$ model\cite{BD} is a minimal extension of $SU(5)$:
the Higgs sector contains supermultiplets $\Sigma\sim 35$ and
$H+\bar{H}\sim 6+\bar{6}$ respectively in adjoint and fundamental 
representations, in analogy to 24 and $5+\bar{5}$ of $SU(5)$.
However, this model drastically differs from the other GUTs 
where the Higgs sector usually consists of two different sets:
one is for the GUT symmetry breaking (e.g. 24-plet in $SU(5)$),
while another containing the Higgs doublets (like $5+\bar{5}$ 
in $SU(5)$) is just for the electroweak symmetry breaking. 
The $SU(6)$ theory has no special superfields for the second
purpose: 35 and $6+\bar 6$ constitute a minimal Higgs content
needed for the local $SU(6)$ symmetry breaking down to 
$SU(3)\times SU(2)\times U(1)$.
As for the MSSM Higgs doublets $\phi_{1,2}$, they are contained 
in $\Sigma$ and $H,\bar{H}$ themselves, as the Goldstone modes 
of the accidental global symmetry $SU(6)_\Sigma \times U(6)_H$ 
related to independent transformations of $\Sigma$ and $H$.
This global symmetry arises if mixing terms of the form $\bar{H}\Sigma H$ 
are suppressed in the Higgs superpotential\cite{BD}. 
%Then 
%$\phi_{1,2}$ being strictly massless in the exact SUSY limit, acquire
%non-zero mass terms (interestingly, including also the $\mu$-term) 
%due to the spontaneous SUSY breaking. 
In particular, if the Higgs superpotential has a `factorized' 
form   
\be{factorized} 
{\cal W}_{\rm Higgs} = {\cal W}(\Sigma) + {\cal W}(H\bar H) 
\ee 
e.g. with ($Y$ is some auxiliary singlet)  
\be{W-su6}
{\cal W}(\Sigma) = M\Sigma^2 + h\Sigma^3 , ~~~~~  
{\cal W}(H\bar H)= \rho Y (\bar{H}H - \Lambda^2)
%+ M_H \bar{H}H + M_Y Y^2 + \xi Y^3
\ee
Then it has an accidental global symmetry 
$SU(6)_{\Sigma}\times U(6)_H$.\footnote{ The global 
$SU(6)_\Sigma\times U(6)_H$ is an accidental symmetry of the Higgs 
{\it superpotential} and not a symmetry of the whole Lagrangian: 
the Yukawa terms and the gauge couplings ($D$-terms) do not respect it.
However, in the exact supersymmetry limit it is effective for the 
field configurations on the vacuum valley where $D=0$. 
Owing to non-renormalization theorem, it cannot be
spoiled by the radiative corrections. }
In the limit of exact SUSY (vanishing $F$ and $D$ terms), 
among the other degenerated vacua there is a following one: 
\be{VEVs}
\langle \Sigma \rangle = V_\Sigma \cdot {\rm diag}(1,1,1,1,-2,-2), 
%\left( \begin{array}{cccccc}
%1 & & & & & \\& 1 & & & & \\ & & 1 & & & \\ & & & 1 & & \\ & & & & -2 & \\
%& & & & & -2 \end{array} \right), 
~~~~ \langle H \rangle =\langle \bar{H} \rangle = V_H\cdot(1,0,0,0,0,0)
%\left( \begin{array}{c}
%V_H\\0\\0\\0\\0\\0 \end{array} \right),~~~
%V_\Sigma=\frac{M}{h}, ~~~ V_H=\La  
\ee
where $V_\Sigma =M/h$ and $V_H=\La $.  
After SUSY breaking, this configuration can indeed be 
a true vacuum state for a proper range of the soft parameters. Then 
$H,\bar{H}$ break $SU(6)$ down to $SU(5)$ while $\Sigma$ breaks
$SU(6)$ down to $SU(4)\times SU(2)\times U(1)$, and both 
channels together lead to the local symmetry breaking down to
$SU(3)\times SU(2)\times U(1)$. 
At the same time, the global symmetry $SU(6)_{\Sigma}\times U(6)_H$
is broken down to $[SU(4)\times SU(2)\times U(1)]_\Sigma \times U(5)_H$.
Most of the Goldstone modes 
%correspond to generators of the broken local $SU(6)$ and they 
are eaten up by the $SU(6)$ gauge superfields through the Higgs 
mechanism. However, 
%since the global symmetry of the ground state exceeds the global one, 
a couple of fragments survive and present in particle spectrum at lower 
energies as the Goldstone superfields. These constitute 
the MSSM Higgs doublets $\phi_{1,2}$ which in terms of the doublet
(anti-doublet) fragments in $\Sigma$ and $H,\bar{H}$ are given as
\be{Higgs}
\phi_2= \cos\eta\phi_{\Sigma} - \sin\eta \phi_H\,,~~~~~
\phi_1= \cos\eta\bar{\phi}_{\Sigma} - \sin\eta\bar{\phi}_{\bar{H}}
\ee
where $\tan\eta=3V_\Sigma/V_H$.  
In the following  we assume that 
$V_H\sim M_6$ is larger than  $V_\Sigma\simeq M_5 \simeq 10^{16}$ GeV, 
as it is motivated by the $SU(5)$ unification of the gauge couplings.  
In this case the doublets $\phi_{1,2}$ dominantly come from $\Sigma$
while in $H,\bar{H}$ they are contained with small
weight $\sim 3V_\Sigma/V_H$.

The scalar fields in $\phi_{1,2}$ then get mass from 
the SSB terms: 
\be{SB_terms}
V_{SB}= (m_S[\theta^2{\cal W}']_F + h.c.) + m_k^2\sum_k|\varphi_k|^2 
\ee
where $\varphi_k$ imply all scalar fields involved, 
$m_k\sim m_S$ are their soft masses, 
and ${\cal W}'$ repeates all structures present in the 
superpotential ${\cal W}$, but is not necessarily proportional to 
${\cal W}$.
Incidentally, the GIFT scenario naturally solves also the
$\mu$-problem. Taking into account the SSB terms 
(\ref{SB_terms}) in minimization of the Higgs potential of $\Sigma$ 
and $H,\bar H$, one observes that the VEV 
$V_\Sigma$ is shifted by an amount of $\sim m_S$ as compared to 
the one of eq. (\ref{VEVs}) calculated in the exact SUSY limit. 
Then substituting these VEVs back in superpotential, this shift 
gives rise to the $\mu \phi_1 \phi_2$ term, with $\mu\sim m_S$. 
Thus, in GIFT scenario the (supersymmetric) $\mu$-term   
emerges as a consequence of the SUSY breaking. On the other hand, 
at the GUT scale 
one obtains for the soft parameters in $L_{\rm Higgs}$: 
\be{bmu}
m_1=m_2=m_\Sigma, ~~~~ B_\mu = \mu^2 + m^2_\Sigma 
\ee 
where $m_\Sigma$ is a soft mass term of $\Sigma$ in (\ref{SB_terms}).  
In fact, this relation is a reflection of the fact that 
since ${\cal W}'$ has the same form as ${\cal W}$, then it also 
has the global global symmetry $SU(6)_{\Sigma}\times U(6)_H$ 
and thus one combination $(\phi_1+\phi_2^\ast)/\sqrt2$ 
of the scalar doublets remains massless as a true Goldstone boson. 

However, in reality the 
SUSY breaking relaxes radiative corrections which lift the vacuum 
degeneracy (mainly due to the large top Yukawa coupling, origin of 
which we will clarify below) and fix the VEVs $v_1$ and $v_2$. 
The effects of radiative corrections leading to the electroweak 
symmetry breaking were studied recently in refs.\cite{CR}.

Thus, the $SU(6)$ model naturally solves both the DT splitting and the
$\mu$ problems. The Higgs doublets $\phi_{1,2}$ remain light, with 
$\mu\sim m_S$, and their triplet partners are superheavy: 
the triplets from $\Sigma$ have masses
$\sim M_5$, while the `Goldstone' triplets from $H, \bar H$ 
are eaten up by the $SU(6)$ gauge superfields
via the Higgs mechanism and thus acquire masses $\sim M_6$.

%\section{ GIFT without FT} 

%Certainly only the $SU(6)$ 
%symmetry itself does allow the mixed term $\bar{H}\Sigma H$
%and it is put to zero by hands, which is essentially a FT. 
%It is not difficult to understand, however, that this term can 
%be forbidden by some additional discrete or abelian symmetry 
%reasons in which case the ..... can emerge automatically as an 
%accidental symmetry. These possibilities will be outlined below. 

In order to built a consistent GIFT model, one has to find some valid
symmetry reasons to forbid the mixed term $\bar{H}\Sigma H$:
otherwise the global symmetry $SU(6)_{\Sigma}\times U(6)_H$
is not accidental and the mixed term 
should be put to zero by hands, which essentially would be a FT.
It is natural to use for this purpose the discrete 
symmetries.\cite{BDM,PLB,BCL}  
%which in principle could emerge in the string theory context.
%In addition, they can provide a proper pattern of the higher 
%order operators inducing the fermion masses. 
%Possible consistent models were suggested in refs.\cite{PLB,BCL}. 
%Below we consider the $SU(6)$ model of ref.\cite{PLB} with the 
%flavour-blind discrete symmetry $Z_3$.  

For example, one can introduce two 35-plets $\Sigma_{1,2}$ and 
impose the following discrete $Z_3$ symmetry\cite{PLB}:  
%\be{Z_3}
$\Sigma_1 \to e^{i\frac{\pi}{3}} \Sigma_1$ and   
$\Sigma_2 \to e^{-i\frac{\pi}{3}} \Sigma_2$, 
%\ee
while $H,\bar H$ and $Y$ are invariant. Then the most general 
renormalizable superpotential of $\Sigma_{1,2}$ has a form: 
\be{W-Sigma}
W(\Sigma)= M\Sigma_1\Sigma_2 + h_1 \Sigma_1^3 + h_2\Sigma_2^3 
\ee 
whereas the mixed terms $H\Sigma_{1,2}\bar H$ are forbidden 
by $Z_3$ symmetry, and thus the Higgs superpotential 
acquires an accidental global symmetry $SU(6)_\Sigma \times U(6)_H$.

However, problem arises if one includes also the higher order 
operators cutoff by $M_P$. Then the terms like 
$\frac{1}{M}H\Sigma_1\Sigma_2 \bar H$ etc., allowed by the $Z_3$ 
symmetry would spoil the accidental global symmetry and 
Higgs bosons will get too large masses. 
Some possibilities avoiding the impact of the Planck scale 
terms were suggested in ref.\cite{BCL}. A very intersting 
way to guarantee the accidental global symmetry at all orders 
in $M_P^{-1}$ was suggested in ref.\cite{DP}  which makes use of 
the anomalous $U(1)$ symmetry. 
%In particular, $H$ and $\bar H$ 
%can both have negative $U(1)$ charges while the charge of $\Sigma$ 
%is zero in which case its superpotential can be taken as .... . 

\section{Fermion masses and mixing } 

%\subsection{ SU(5):  } 

In the minimal SUSY $SU(5)$ model the quarks and leptons 
of each family fit into the multiplets 
%\be{su5}
$\bar{5}_i=(d^c_i + l_i)$ and $10_i=(u^c_i + q_i + e^c_i)$,   
i=1,2,3.  
%\ee 
The fermion masses are induced by the Yukawa terms: 
\be{su5-Yuk} 
{\cal W}_{\rm Yuk}= 
\la^u_{ij} 10_i H 10_j\, + \, \la^d_{ij} 10_i \bar{H} \bar{5}_j  
%~~~~~~~~~ 
%{\cal W}_\nu= \frac{\la^{\nu}_{ij}}{M_P} (\bar 5_i H)(H \bar 5_j)
\ee 
At the GUT scale $M_5$ these terms reduce to the MSSM couplings 
(\ref{MSSM}) with $\hat{\la}^e_{ij}=\hat{\la}^{d}_{ji}$, and hence 
$\la_{d,s,b}=\la_{e,\mu,\tau}$. 
The $\la_b=\la_\tau$ unification\cite{BEGN} is a definite success. 
After accounting for the RG running it translates 
into relation for physical masses $m_b/m_\tau \sim 3$, while 
the more precise comparison of $m_b$ and $m_\tau$ within the 
present experimental uncertainties implies that $\la_t$ 
should be rather large ($\geq 1$), in which case 
the top mass is fixed by its infrared limit 
%\be{Top}
$M_t= (190-210)\sin\beta~ {\rm GeV}= 140-210~ {\rm GeV}$. 
%\ee 
Thus, the minimal SUSY $SU(5)$ model explains the principal origin 
of the bottom quark mass and nicely links it to the large value of 
the top mass. 
%within the range indicated by the present data. 

Unfortunately, the other predictions 
$\la_s=\la_\mu$ and $\la_d=\la_e$ are wrong: experimentally 
$m_s/m_d\simeq 20$ while $m_\mu/m_e= 200$. 
In addition, there is no explanation neither for the 
fermion mass hierarchy nor for the CKM mixing pattern: 
the Yukawa matrices $\la^u$ and $\la^d$ remain arbitrary 
and there is no reason neither for hierarchy of 
their eigenvalues nor for their allignment.  
Therefore, one is forced to go beyond the minimal $SU(5)$ model and 
implement new ideas that could shed some more light on the 
origin of fermion masses and mixing\cite{ICTP}. 

%\subsection{ Fermion Masses and Mixing in SUSY $SU(6)$ }
%\vspace{-0.35cm}

In the GIFT $SU(6)$ model the situation is drastically different. 
The fermion sector of the $SU(6)$ theory consists of three families 
$(\bar 6 + \bar 6' + 15)_i$, $i=1,2,3$, and one 20-plet 
(since 20 is a pseudo-real representation, its mass term is 
vanishing)\cite{BDSBH,PLB}. In terms of the $SU(5)$ subgroup the 
fermions under consideration read 
\beqn{fragments}
& & 20=10 + \ov{10} = (q+u^c+e^c)_{10}+(Q^c+U+E)_{\ov{10}} \nonumber \\
& & 15_i=(10+5)_i = (q_i+u^c_i+e^c_i)_{10} +
(D_i + L^c_i)_5  \nonumber \\
& & \bar{6}_i=(\bar{5}+1)_i = (d^c_i + l_i)_{\bar{5}} + n_i , ~~~~ 
 \bar{6}_i^\prime=(\bar{5}+1)_i^\prime =
(D^c_i +L_i)_{\bar{5}^\prime} + n_i^\prime\,,
%~~~~~~~~~~~~~~~~~~~~~i=1,2,3
\eeqn
Only the following renormalizable Yukawa terms allowed by 
the $SU(6)$ symmetry: 
\be{Yukawa} 
{\cal W}_{Yuk} = G\,20 \Sigma 20 \,+ \,\Gamma\,20 H 15_3\,  + \,
\Gamma_{ij} 15_i \bar{H} \bar{6}^\prime_j\,, ~~~~~~~~i,j=1,2,3
\ee 
(all coupling constants are assumed to be $\sim 1$). 
Without loss of generality, the basis of 15-plets is chosen  
so that only the $15_3$ state couples 20-plet in (\ref{Yukawa}).
Also, among six $\bar{6}$-plets one can always choose three of them
(denoted in eq. (\ref{Yukawa}) as $\bar{6}^\prime_{1,2,3}$) which couple
$15_{1,2,3}$ while the other three states $\bar{6}_{1,2,3}$ have
no Yukawa couplings.

Already at the scale $V_H$ of the gauge symmetry breaking
$SU(6) \to SU(5)$ the fermion content reduces to the one of the 
minimal $SU(5)$. 
%In the spirit of survival hypothesis\cite{surv}, 
The extra vector-like fermions $\ov{10}+10_3$ and $(5+\bar{5}')_{1,2,3}$,
get $\sim V_H$ masses from couplings (\ref{Yukawa}) 
%\be{heavymass}
%\Gamma\,V_H\,\ov{10}\,10_3\, +
%\,\Gamma_{ij}V_H\,5_i\,\bar{5}_j^\prime\, +
%\,G\, V_\Sigma \,(U\,u^c - 2 E\,e^c) \,,
%\ee
and thereby the light states remain as
$\bar{5}_{1,2,3}$, $10_{1,2},10$ and singlets $n_i,n'_i$.  
%(we neglect $\sim \eps_\Sigma$ mixing between the 
%$u^c - u^c_3$ and $e^c - e^c_3$ states). 
 
In the $SU(6)$ model the Yukawa couplings of the Higgs doublets 
to fermions are very peculiar:  
%leading to new 
%possibilities towards understanding of the flavour structure. 
should the Yukawa terms also respect the global 
$SU(6)_\Sigma \times U(6)_H$ symmetry, then $\phi_{1,2}$ 
being the Goldstone modes would have the {\em vanishing} Yukawa 
couplings to all fermions which remain massless after the GUT 
symmetry breaking down to the MSSM, that are ordinary 
quarks and leptons. Thus, the couplings relevant for fermion 
masses have to {\em explicitly} violate $SU(6)_\Sigma \times U(6)_H$. 
This leads to striking possibilities to understand 
the key features of the fermion mass and mixing 
without invoking the additional symmetry arguments.
In particular, the {\em only} fermion which can get $\sim 100$ GeV 
mass through the renormalizable Yukawa coupling is the top quark, 
while other fermion masses can emerge only through the higher order 
operators and thus are suppressed by powers of the Planck scale $M_P$.
Therefore, the observed hierarchy of fermion masses 
and mixings can be naturally explained in terms of small ratios
$\eps_\Sigma=V_\Sigma/V_H$ and $\eps_H=V_H/M_P$.\cite{BDSBH,PLB,BCL}  

%In the following we assume that all coupling constants in the Higgs 
%as well as in the Yukawa sectors are of the order of 1. 
%(For comparison, the gauge coupling constant
%at the GUT scale is $g_X\simeq 0.7$.)  

Indeed, 
the couplings of 20-plet in (\ref{Yukawa}) explicitly
violate the global $SU(6)_\Sigma\times U(6)_H$ symmetry. Hence,
the up-type quark from 20 (to be identified as top)
has {\em non-vanishing} coupling with the Higgs doublet $\phi_2$.
As far as $V_H\gg V_\Sigma$, it essentially emerges
from $20\Sigma 20 \to  q u^c \phi_2.\,$
Thus, {\em only} the top quark can have $\sim 100\,$GeV
mass due to the large Yukawa constant $G$: $\lambda_t\sim 1$.

All other fermion masses can be induced only from the higher order 
operators scaled by the inverse powers of $M_P$. These operators 
can emerge effectively by integrating out some heavy fermion 
states.\cite{FN}  
For example, the following $d=6$ operator is responsible for 
the $b$ and $\tau$ masses: 
%\footnote{The terms like $15 \bar H (\Sigma^2 \bar 6)$ or 
%$15 \bar H \bar 6\cdot {\rm Tr}(\Sigma^2)$ do not violate 
%the global $SU(6)_\Sigma\times U(6)_H$ symmetry and are 
%therefore irrelevant. }
\be{B}
{\cal B} = \frac{B}{M_P^2}\, 20\bar{H}(\Sigma\bar{H})\bar{6}_i\, 
\ee 
%Operator ${\cal B}$ gives rise to the $b$ quark and $\tau$ lepton masses. 
At the MSSM level it reduces to the Yukawa
couplings $\eps_H^2 B (q d^c_3  + e^c l_3)\phi_1$.
Hence, though $b$ and $\tau$ belong to the 20-plet as well as 
$t$, their Yukawa constants are by factor $\sim \eps_H^2$ smaller 
than $\lambda_t$. In addition, the $b-\tau$ Yukawa constants are 
automatically unified at the GUT scale --- 
up to $\sim \eps_\Sigma^2$ corrections, $\la_b=\la_\tau$.    
%due to the mixing of $e^c$ and $e^c_3$ states.

The second family fermions can get masses from the operators: 
\beqn{dim6}
%&{\cal B} = \frac{1}{M_P^2}\, 20\bar{H}(\Sigma\bar{H})\bar{6}_3\,,
& {\cal C} = \frac{ C_{ij} }{M_P^2}\, 15_i H (\Sigma H) 15_j , ~~~~ 
{\cal C}' = \frac{ C_{i} }{M_P}\, 20 \Sigma H 15_i
\nonumber \\ 
%\be{dim4_smu}
& {\cal S} = \frac{{S}_{ik}}{M_P^2}\, 15_i(\Sigma^2\bar{H})\bar{6}_k , 
~~~~~ {\cal S}' =
\frac{{S}'_{ik}}{M_P^2}\,15_i(\Sigma\bar{H})(\Sigma\bar{6}_k)
\eeqn
($SU(6)$ indices are contracted so that combinations 
in the parentheses transform as effective $\bar 6$ or $6$).
Since the $\bar 6'$ and $15_3$ states already have $\sim V_H$ masses, 
these operators are relevant only for the light states in 
$20$, $15_{1,2}$ and $\bar{6}_{1,2,3}$. 
One can always redefine the basis of $\bar{6}$-plets so that only 
the $\bar{6}_3$ couples 20 in eq. (\ref{dim6}). In addition, we 
assume that constants $B$, $C_{ij}$, etc. all are order 1 as well 
as the constants in (\ref{Yukawa}).

%Operator ${\cal C}$ contributes the up quark Yukawa constants 
%of the first and second families, as 
%$\lambda_{ij}^u=\eps_H^2 C_{ij}$ ($i,j=1,2$). 
%As for the operators ${\cal S}$ and ${\cal D}$, they induce the Yukawa 
%constants of the down quarks and charged leptons respectively as 

This allows to explain the observed hierarchy of 
fermion masses and mixing in terms of two small parameters, 
$\eps_\Sigma,\eps_H \sim 0.1$.\footnote{
As far as  the scale
$M_5\simeq 10^{16}\,$GeV is fixed by the $SU(5)$ unification
of the gauge couplings, this corresponds to $M_6\sim 10^{17}\,$GeV 
and $M_P \sim 10^{18}\,$GeV, so that $M_P$ 
is indeed close to the string or Planck scale.
The relation $M_6\sim \sqrt{M_5 M_P}$ could naturally emerge in 
the context of the models discussed in refs.\cite{PLB,BCL,DP}. 
}
More details on the fermion mass structures 
can be found in refs.\cite{BDSBH,PLB,BCL}. 
%even without appealing to any horizontal symmetry reasons. 
% provided that the scales $M_P$, $V_H$ and
%$V_\Sigma$ are related as $V_\Sigma/V_H \sim V_H/M_P \sim 0.1$.
In the context of these one can obtain the following hierarchy 
for the Yukawa coupling eigenvalues at the GUT scale 
\beqn{Yuk_pattern}
\lambda_t\sim 1, ~~~ \lambda_b=\lambda_\tau\sim \la_c \sim \eps^2, 
~~~ \lambda_c\sim \eps^2,  
~~~ \la_s,\lambda_\mu \sim \eps^3,   
~~~ \lambda_u,\la_d,\lambda_e \sim \eps^5 
\eeqn  
%\beqn{Yukawa_pattern}
% 3^{rd} ~{\rm family}:~~~~ & \lambda_t\sim 1,
%&\lambda_\tau=\lambda_b\sim \eps_H^2     \nonumber \\
% 2^{nd} ~{\rm family}:~~~~ & \lambda_c\sim \eps_H^2,  
%&\lambda_\mu= 5\lambda_s \sim \eps_\Sigma\eps_H^2     \nonumber \\
% 1^{st} ~{\rm family}:~~~~ & \lambda_u\sim \eps_\Sigma^2\eps_H^4, 
%&\lambda_e= \frac{5}{8} \lambda_d \sim \eps_\Sigma^2\eps_H^3 
%\eeqn  
and for the CKM angles 
%(notice different parametrization from that of eq. (\ref{CKM})): 
\be{CKM-new}
%V_{\rm CKM}\approx
%\matr{ 1 }{ s_{12} }{ s_{12}s_{23} - s_{13}e^{-i\delta} }
%{ -s_{12} }{ 1 }{ s_{23} + s_{12}s_{13}e^{-i\delta} }
%{ s_{13}e^{i\delta} }{ -s_{23} }{ 1 }, ~~~~
s_{12}\sim 1,~~~~ 
s_{23}\sim \eps, ~~~~ 
%\frac{\lambda_s}{\lambda_b}, ~~
s_{13}\sim \eps^2 
%\frac{\lambda_d}{\lambda_b}
\ee
%where the CP-phase $\delta$ arises due to the complex Yukawa 
%constants in the theory.  
%
The lowest dimension operator relevant for the neutrino masses 
is 
%\be{N}
${\cal N} = \frac{N_{kl} }{M_P^2}\, \bar{6}_k H \Sigma H \bar{6}_l$.
%\ee
For $\eps_\Sigma,\eps_H\sim 0.1$ it induces the small Majorana masses 
$m_\nu \sim v^2/M_P \sim 10^{-5}\,$eV. 
This mass range (which as a matter of fact 
is also favoured by the minimal $SU(5)$ model\cite{BEG})  
is just what is needed 
%in constants  $N_{kl}\sim 1$, one expects that neutrino mixing 
%angles are $O(1)$. 
%Thus the predictions for the neutrino oscillation 
%parameters are in the range (\ref{JS}), needed 
for the long wavelength ``just-so'' oscillation 
solution to the solar neutrino problem.

\section{discussion} 

Let us conclude with a brief summary of the simplest SUSY GUTs 
and appropriate mechanisms for the 2/3 splitting. 
%MDM in $SU(5)$, MVM in $SO(10)$ and GIFT mechanism in $SU(6)$. 
Their main features are given in Table 1.

\begin{table}[h]
\tcaption{
Summary of the known solutions to the 2/3 splitting problem 
in various SUSY GUTs: MDM in $SU(5)$, MVM in $SO(10)$ and 
GIFT in $SU(6)$, with the minimal multiplet contents required 
for realization of these models.
The lowest Kac-Moody levels at which these 
models can be in principle embeddable in the string theory are
also shown. 
}\label{tab:exp}
\small
\begin{tabular}{||c|c|c|c|c||}\hline\hline
{} & {} & {} & {} & {} \\
GUT & Higgses & Fermions & $\mu$ and $B_\mu$ & Yukawas \\
{} & {} & {} & {} & {} \\
\hline
{} & {} & {} & {} & {} \\
$SU(5)$ & $5 + \bar5 + 75$ & $(\bar5 +10)_{i}$ & $?$ 
& $\la_b=\la_\tau$ \\
($k=4$) & $+50+\ov{50}$ & {} & {} & $(\tanb=?)$ \\
{} & {} & {} & {} & {} \\ 
\hline
{} &{} &{} &{} & {} \\
$SO(10)$  & $10+10'+ 45_{1,2,3}$ & $16_i$ & $?$ 
& $\la_b=\la_\tau=\la_t\sim 1$  \\ 
($k=2$) & $+54 +16 + \ov{16}$ &{} &{} & $(\tanb\sim 100)$ \\  %[-37pt]
{} & {} & {} & {} & {}\\
\hline 
{} & {} & {} & {} & {} \\
$SU(6)$ & $6+\bar{6}+35$ & $27_i+20$ & 
$\mu\sim m_S$ & 
$\la_t \sim 1$, ~~ $\la_b=\la_\tau \sim \eps^2$ \\
($k=2$)  &{} &{} & $B_\mu=\mu^2 + m^2_{1,2}$ & $\la_c \sim \eps^2$, 
~~ $\la_{s,\mu}\sim \eps^3$ \\ 
%[-37pt]
{} &{} & {} & {} & $(\tanb \sim 1)$ \\
%[24pt]
\hline\hline
\end{tabular} 
\end{table}

The only known solution of the 2/3 splitting problem 
in SUSY $SU(5)$ is based on MDM\cite{MDM} or its variations 
(see ref.\cite{IMDM} and references therein). 
It requires rather complicated Higgs sector involving  
large representations which in the string theory context 
%the representations 75, 50 and $\ov{50}$ 
can be allowed only at the Kac-Moody level $k_5\geq 4$. 
The main problem of the MDM is that the extra states 50 and $\ov{50}$  
with masses $\sim M_5$ contribute the RG running 
up of the $SU(5)$ gauge constant and drive it out the perturbative 
regime well below the Planck scale $M_P$. Therefore, it is hard 
to imagine how such a model can be embedded in a string theory. 
Situation could be saved if the mass of 50-plets is $\sim M_P$.  
However, 
in this case the mass of the triplets $T,\bar T$ from $H,\bar H$ 
is $M_3\sim M_5^2/M_P\sim 10^{14}$ GeV which would cause 
catastrophically fast proton decay via $d=5$ operators. 

The MVM solution\cite{DiWi} operative in SUSY $SO(10)$ is also 
rather cumbersome and requires complicated Higgs sector. 
It also can be justified by symmetry reasons\cite{IMVM}. 

Neither the MDM nor the MVM solutions for 2/3 splitting 
do not touch the $\mu$-problem, 
and for its solution some additional ideas should be incorporated. 
In addition, the $SU(5)$ or $SO(10)$ models themselves  
give no understanding to the fermion mass hierarchy, though 
by involving the horizontal symmetry concept one can obtain 
the predictive mass textures.\cite{ADHRS}  

The $SU(6)$ model\cite{BD} seems to be most favourable,  
and The GIFT mechanism can be justified by symmetry 
reasons\cite{BCL,DP}. The Higgs sector of this theory 
is simple: it contains only the adjoint and fundamental 
representations. The fermion sector is also simple -- 
three families of supermultiplets 
$(\bar 6 + \bar 6' +15)_i=27_i$, $i=1,2,3$, which in fact 
constitute the 27-plets of $E_6$ group, and one 20-plet. 
Therefore, there is no difficulty in maintaining the 
perturbative regime for the gauge coupling up to the scale $M_P$. 

The $SU(6)$ model 
provides a simoultaneous solution to both 2/3 splitting and 
$\mu$ problems, leads to prediction for the soft parameters 
$m_{1,2}$ and $B_\mu$ pattern at the GUT scale, and, in addition, 
explains many features of the fermion mass spectrum and mixing 
by internal reasons, without involving additional ideas.

%{\bf References} 
%\vspace{0.7cm}

\end{document}